\documentclass[aip,cha,reprint,numerical]{revtex4-1}
\usepackage{amsmath}
\usepackage{amssymb}
\usepackage{graphicx}
\usepackage{dcolumn}
\usepackage{bm}
\usepackage{natbib}		
\usepackage{color}

\newcommand{\pd}[2]{\displaystyle\frac{\partial #1}{\partial #2}}

\newcommand {\e} {\varepsilon}
\newcommand {\vp} {\varphi}

\def\w{\omega}

\begin{document}
\title{Numerical phase reduction beyond the first order approximation}

\author{Michael Rosenblum}
\affiliation{Institute of Physics and Astronomy, University of Potsdam, 
Karl-Liebknecht-Str. 24/25, 14476 Potsdam-Golm, Germany}
\affiliation{Control Theory Department, Institute of Information Technologies,
Mathematics and Mechanics, Lobachevsky University Nizhny Novgorod, Russia}
\author{Arkady Pikovsky}
\affiliation{Institute of Physics and Astronomy, University of Potsdam, 
Karl-Liebknecht-Str. 24/25, 14476 Potsdam-Golm, Germany}
\affiliation{Control Theory Department, Institute of Information Technologies, 
Mathematics and Mechanics, Lobachevsky University Nizhny Novgorod, Russia}

\date{\today}

\begin{abstract}
We develop a numerical approach to reconstruct the phase dynamics of driven or 
coupled self-sustained oscillators. Employing a simple algorithm for 
computation of the phase of a perturbed system, we construct numerically the  
equation for the evolution of the phase. 
Our  simulations demonstrate that the description of the dynamics solely by phase 
variables can be valid for rather strong coupling strengths and large deviations from the 
limit cycle. Coupling functions depend crucially on the coupling and are generally
non-decomposable in phase response and forcing terms.
We also discuss limitations of the approach.  
\end{abstract}

\maketitle

\begin{quotation}
It is widely accepted that the dynamics of weakly interacting limit 
cycle oscillators can be fully described by their phases, while the 
 amplitudes can be considered as enslaved and thus irrelevant. 
This idea is commonly used both in theoretical and experimental
studies. While in the latter case the phase dynamics is inferred from data, 
in the former one, the corresponding equation shall be derived by a 
perturbation technique. However, the existing theory provides
only a recipe for computation of the first order approximation. 
In spite of numerous recent efforts, phase dynamics description 
beyond the limit case of weak coupling is still lacking. 
Here we provide a numerical approach for the phase dynamics reduction beyond
the first order approximation.
\end{quotation}

\section{Introduction}
Phase dynamics approximation plays an important role in the analysis of coupled
self-sustained oscillators 
\cite{Winfree-80,Kuramoto-84,Hoppensteadt-Izhikevich-97,Pikovsky-Rosenblum-Kurths-01,%
Ermentrout-Terman-10,Monga_Wilson-Matchen-Moehlis-18}. 
Reduction to the phase description essentially decreases
the dimension of the problem and in some cases provides an analytical solution.
Well-known examples of the solvable phase dynamics are the Adler 
equation \cite{Adler-46} for a periodically 
driven oscillator and the Kuramoto model of (infinitely) many globally 
coupled systems \cite{Kuramoto-84}. 
However, a practical application of the theory and a derivation of the phase dynamics equation 
from equations formulated in state variables
for an arbitrary driven or coupled oscillator remains an unsolved problem: this 
derivation requires an analytical expression for isochrons \cite{Guckenheimer-75} 
or for the phase response curves, 
and therefore can 
be accomplished analytically only in exceptional cases, e.g., for the Stuart-Landau system. 
Moreover, this equation can be obtained only in the first approximation,
as the first-order term of the perturbative expansion with respect 
to small coupling parameter \cite{Kuramoto-84}. 
In spite of recent attempts (see a discussion in 
Refs.~\onlinecite{Monga_Wilson-Matchen-Moehlis-18,Wilson-Ermentrout-18}), there exist no 
practical algorithms for inferring phase description in a general case when a perturbed
trajectory is not very close to the unperturbed limit cycle.

In this communication we propose an efficient numerical approach which provides
the desired phase dynamics equation. The evolution of the phase is described by a coupling 
function that is either defined on a fine grid or represented by 
a finite Fourier series.
Next, we use this approach to study the phase dynamics for strongly driven 
systems, where we cannot expect validity of the first order approximation.
For two different examples we reconstruct the second- 
and the third-order terms of the coupling functions, and verify that this
description works with a good precision.

We start by introducing our notations and recalling main ideas of the phase 
reduction theory. Let the autonomous system be described by 
$\mathbf{\dot X=F(X)}$, where $\mathbf{X}$ is the state vector in $n$-dimensional 
phase space, $n\ge 2$. This equation defines the flow $(\mathbf{X},t)\to S^t\mathbf{X}$.
Our goal is to find such a function of the phase space $\Phi(\mathbf{X})$, mapping
the state of the system onto a unit circle, that allows to attribute to every
trajectory $\mathbf{X}(t)$ the corresponding phase evolution $\varphi(t)=\Phi(\mathbf{X}(t))$.
Depending on the context, we will either treat $\varphi$ as an wrapped phase (points on a unit circle),
or as unwrapped phase with $-\infty<\varphi<\infty$. 

If  the system possesses a stable limit cycle 
$\mathbf{\overline{X}}$ with period $T$, then the phase definition $\Phi(\mathbf{X})$
can be constructed for all states $\mathbf{X}$ in the basin of attraction of $\mathbf{\overline{X}}$.
First, the phase is easily introduced on $\mathbf{\overline{X}}$. One assigns zero value of $\vp$ to an arbitrary point 
$\mathbf{X_0}$ on the limit cycle. Then, for $0\leq t<T$, the phase on the limit cycle is defined as
$\Phi(S^t\mathbf{X_0})=\omega t$, with $\w=2\pi/T$.  

Extension of this definition to the whole basin of attraction
of the cycle  $\mathbf{\overline{X}}$ is based on the isochrons~\cite{Guckenheimer-75}.
For a point $\mathbf{X}$ in the basin of attraction of $\mathbf{\overline{X}}$, the phase 
is defined as $\Phi(\mathbf{X})=\Phi(\mathbf{X_*})$, where
$\mathbf{X_*}=\lim_{m\to\infty}S^{mT}\mathbf{X}$ ($m=0,1,2\ldots$); 
obviously this limiting point belongs to the limit cycle, 
$\mathbf{X_*}\in \mathbf{\overline{X}}$. 
In other words, isochrons, as the sets of equal phases,
are stable manifolds of the corresponding points on the limit cycle under
the action of the stroboscopic map $\mathbf{X}\to S^T \mathbf{X}$.

Knowledge of isochrons, i.e. of the relation $\vp=\Phi(\mathbf{X})$ in a
vicinity of the limit cycle, allows one to obtain
phase equations for a \textit{perturbed} system 
\begin{equation}
\mathbf{\dot X=F(X)}+\e \mathbf{p}(\mathbf{X},t)\;,
\label{pert}
\end{equation}
where $\mathbf{p}$ describes external forcing or coupling with the strength $\e$. 
Then, for small $\e$, exploiting the perturbation approach,
\textit{using the unperturbed definition of the phase},
and neglecting deviations from the limit cycle 
(see \cite{Kuramoto-84,Pikovsky-Rosenblum-Kurths-01} for details), 
one obtains 
\begin{equation}
\dot\vp=\omega+\e \sum_k \pd{\Phi(\mathbf{\overline{X}}(\vp))}{X_k}p_k(\mathbf{\overline{X}}(\vp),t)=
\omega+\e Q_1(\vp,t)\;.
\label{eq1}
\end{equation}
where $\e Q_1(\vp,t)$ is called the coupling function and the subscript emphasizes 
that the latter is obtained in the first order approximation in $\e$. 
In typical cases, e.g., for periodic external forcing or coupling to another 
limit cycle oscillator, $\mathbf{p}$ can be parameterized by its phase, $\psi$ 
(it means that we can write $\mathbf{p}(\mathbf{X},\psi)$),
and then Eq.~(\ref{eq1})
can be written as
\begin{equation}
 \dot\vp=\omega+\e Q_1(\vp,\psi)\;.
\label{eq2}
\end{equation}

Generally, the perturbation approach implies that the dynamics can 
be represented by a power series in $\e$:
\begin{equation}
 \dot\vp=\omega+\e Q_1(\vp,\psi)+\e^2 Q_2(\vp,\psi)+\ldots =\omega+Q(\vp,\psi,\e)  \;.
\label{eq3}
\end{equation}
One can expect that this description is valid not only for small coupling, but as long 
as the dynamics of the driven system occurs on a smooth attractive torus.
Although the methods for computation of the high-order terms are not known, 
the function $Q$, and the corresponding contributions $Q_k$, $k=1,2,\ldots$, 
can be obtained numerically, as is demonstrated below.

To complete the introduction we mention that if only one component of the force 
in~\eqref{pert} is present, and this force
does not depend on the state of the system, i.e., $\mathbf{p}(\mathbf{X},t)=p(t)$,
then, according to~\eqref{eq1}, the coupling function in the first approximation
can be represented as a product
\begin{equation}
Q_1(\vp,t)=Z(\vp)p(t)\;,
\label{eq:wf}
\end{equation}
where $Z(\vp)$ is the phase sensitivity function, also called the phase response curve (PRC).
This representation is often called the Winfree form.

\section{Numerical computation of coupling functions}
In this Section we describe the numerical aspects of the technique and 
present our algorithm step-by-step. 

\subsection{Inference of the coupling function}
\begin{enumerate}
\item The first, preparatory step is to compute period $T$ of the autonomous oscillation 
\footnote{We use the known Henon-trick~\cite{Henon-82} for precise 
determination of the times when the trajectory returns to $\mathbf{X_0}$.}
and to choose a zero-phase point $\mathbf{X_0}\in\mathbf{\overline{X}}$
on the limit cycle.
\item The next step is to obtain the time series $\mathbf{X}(t_k)$. 
This is accomplished by virtue of numerical integration 
of ODEs~\eqref{pert}, e.g. by a Runge-Kutta method.
\item For each point $\mathbf{X}(t_k)$, the phase $\vp(t_k)=\Phi(\mathbf{X}(t_k))$ is 
calculated. The key issue here is
numerical evaluation of the function $\Phi(\mathbf{\tilde X})$, 
which should provide the phase for any state $\mathbf{\tilde X}$. 
According to the theory outlined above, this function is determined solely by the autonomous dynamics. Then, for each $\mathbf{X}(t_k)$ we proceed as follows.
\begin{enumerate}
\item We integrate the autonomous equations   $\mathbf{\dot Y=F(Y)}$, taking 
$\mathbf{Y}(0)=\mathbf{X}(t_k)$ for the initial condition. 
The integration time is chosen to be $NT$,
where an integer $N$ shall be taken sufficiently large so that we can assume
that the trajectory is already attracted to the limit cycle. (In fact, $N$ is the 
only parameter of our procedure, it crucially depends on the second largest
multiplier of the limit cycle.) 
\item Next, we compute the phase of the 
point $\mathbf{Y}(NT)$. For this goal we determine the time $\tau$ required for the 
trajectory started from $\mathbf{Y}(NT)\in\mathbf{\overline{X}}$ to achieve $\mathbf{X_0}$. 
This provides
$\Phi(\mathbf{X}(t_k))=\Phi(\mathbf{Y}(NT))=2\pi(T-\tau)/T$.
\end{enumerate}
We emphasize that steps 3a,b are not required if the function 
$\Phi(\mathbf{\tilde X})$ is known explicitly, like
for the Stuart-Landau oscillator\cite{Pikovsky-Rosenblum-Kurths-01}. 
However, even in this exceptional case, further numerical
steps for determination of the coupling function beyond the first order 
approximation appear to be unavoidable.
\item Next, we evaluate numerically $\dot\vp_k=\dot\vp(t_k)$.
Practically, we use the Savitzky-Golay smoothing filter
for numerical differentiation of $\vp_k$, see Refs.~\onlinecite{Kralemann_et_al-13,Rosenblum-Pikovsky-18}
for details.
\item As a result of the previous steps we obtain three scalar time series 
$\{\dot\vp_k,\vp_k,\psi_k\}$, where $\psi_k=\omega t_k$. Now we use these series to 
fit the r.h.s. of Eq.~(\ref{eq3}) and thus obtain $Q(\vp,\psi)$ on the 
torus $0\leq \vp,\psi<2\pi$. 
Practically, we obtain $Q$ on a grid $100\times 100$ using kernel density function 
estimation \cite{Chen_KernelTutorium-17}; alternatively it can be 
obtained as a finite double Fourier series.
Notice that the fit is possible if the 
systems remains in a quasiperiodic state so that the trajectory densely fills
the torus.
 \end{enumerate}

 Our final remark is related to software used. For the steps 4 and 5 we used 
functions \verb|co_phidot1| and \verb|co_kcplfct1| 
from our open-source Matlab toolbox DAMOCO, available at 
www.stat.physik.uni-potsdam.de/$\sim$mros/damoco2.html. 
An example of C code for computation of phases (steps 1-3) is available upon request.
 
\subsection{Inference of the coupling function in the Winfree form}
As follows from Eq.~(\ref{eq:wf}), for a known force $p(t)$, determination 
of the coupling function reduces to finding PRC. 
For infinitely weak coupling the latter can be obtained from equations of 
the autonomous system by solving an adjoint problem 
\cite{Monga_Wilson-Matchen-Moehlis-18,Ermentrout-Kopell-91}. 
Since our goal here is to explore the strong coupling case and to 
check whether the representation via a product still holds, 
we compute PRC for different values of $\e$ from  
Eq.~(\ref{eq:wf}). Since $\{\dot\vp_k,\vp_k,\psi_k\}$ are already computed,
this computation can be accomplished by expanding PRC in a Fourier series and obtaining the
coefficients of this series via linear fit.

\subsection{Decomposition in powers of the coupling strength $\e$}
\label{secdecomp}
Here we shortly discuss how a general coupling function
$Q(\vp,\psi,\e)$ can be decomposed in a power series of $\e$, cf.~Eq.~(\ref{eq3}).
In our simulations we restrict the order of the series to three; hence, we are looking
for a third-degree polynomial representation of $Q(\varphi,\psi,\e)$ and determine 
$Q_{1,2,3}$.
For this goal it suffices to compute $Q(\varphi,\psi,\e)$ for at least three different 
values of coupling $\e$. Practically, in our examples shown below, we use more than 10 
values of $\e$. 
Then, with account of the condition $Q(\vp,\psi,0)=0$,
the polynomial $\e$-dependence can be found by least mean squares fit.
Notice that this procedure can be performed point-wisely for coupling functions given on a grid
(below we use this option), or for each Fourier harmonics for the Fourier representation of $Q$.
As a result, the partial coupling functions $Q_{1,2,3}$ defined in~\eqref{eq3} are obtained.

\section{Results} 
In order to illustrate the advantages and limitations of our approach we 
analyze the phase dynamics of two simple paradigmatic models. 
 
\subsection{The Rayleigh oscillator}
In the first example we consider the 
harmonically forced Rayleigh oscillator
\begin{equation}
\ddot x-\mu(1-\dot x^2)\dot x+x=\e \cos(\nu t)\;.
\label{ray}
\end{equation}
The nonlinearity parameter
is $\mu=4$; for this value the limit cycle is strongly stable 
(its transversal Lyapunov exponent is $\approx -5.62$, what for 
the period of the cycle $\approx 10.2$ yields a
multiplier $\approx 10^{-24}$.)
Notice that in this particular case the perturbation is scalar and therefore 
the first-order coupling function can 
be written in the Winfree form~\eqref{eq:wf} with $p(t)=\cos\psi=\cos(\nu t)$. 

We fix the frequency of the forcing at $\nu=0.8$ and increase the amplitude
from a small value $\e=0.01$ (for which we can expect that the first-order phase 
approximation is valid) up to $\e=0.55$, which is not small at all:
this forcing is comparable to the amplitude of $x(t)$. 
Notice that for larger values of $\e$ the system gets locked to the force and 
determination of the coupling function with our method becomes impossible.

\begin{figure}[thb!]
\centerline{\includegraphics[width=0.5\textwidth]{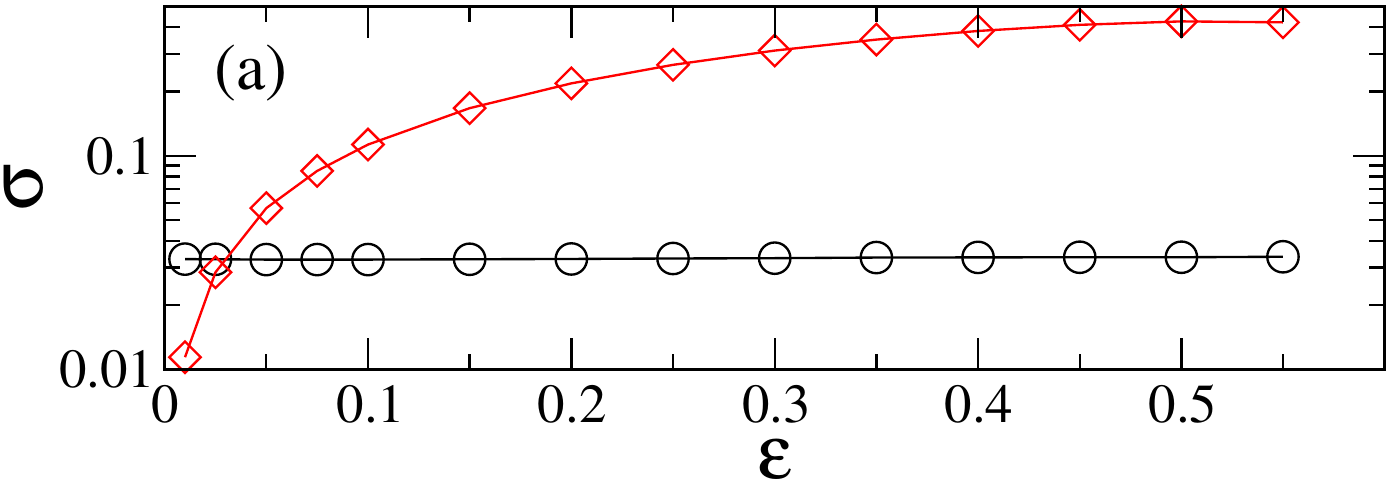}}
\centerline{\includegraphics[width=0.5\textwidth]{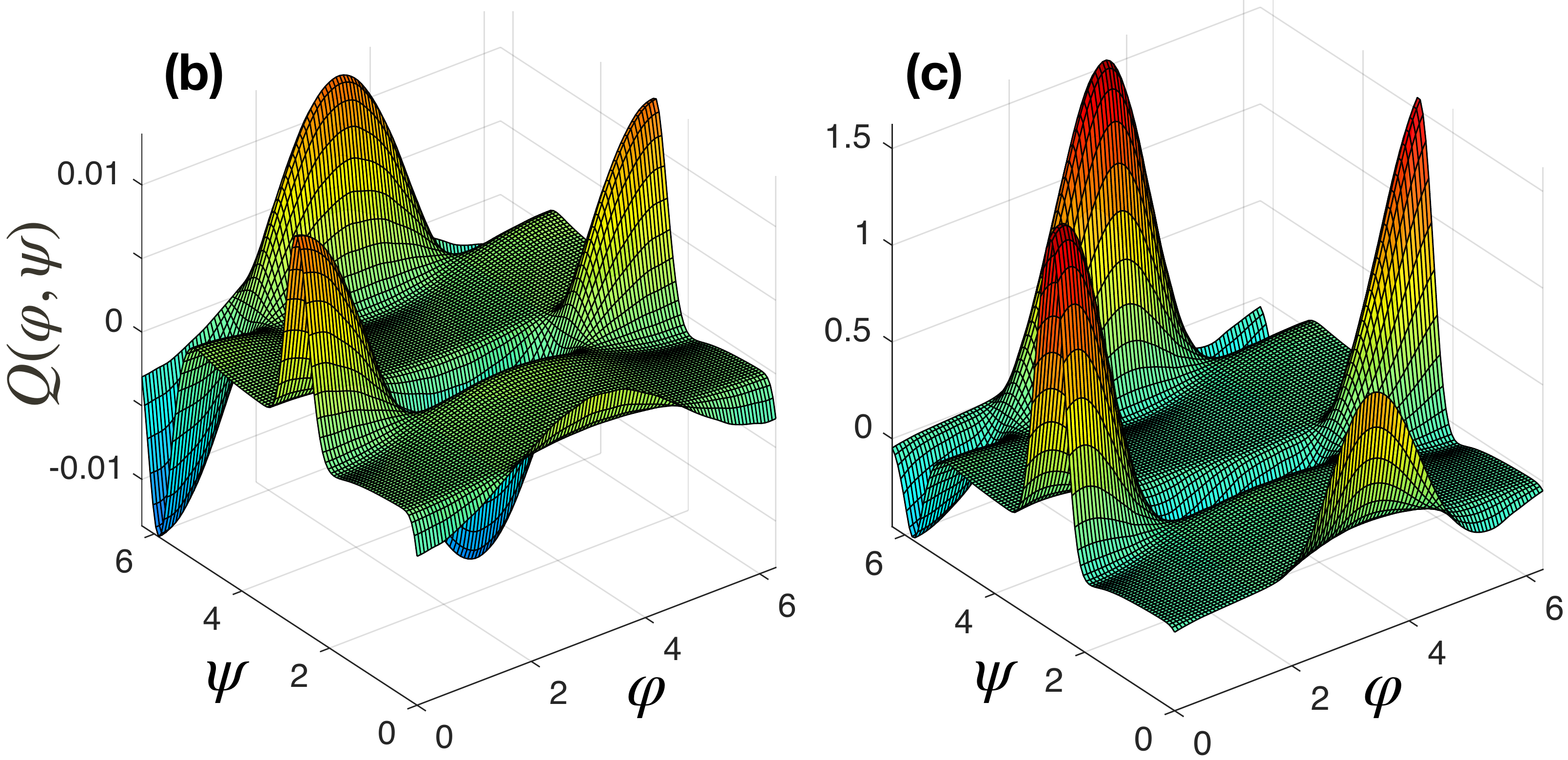}}
\caption{(a) Quality of phase dynamics description in terms of the coupling function
(see Eq.~(\ref{eq3}); circles), and in terms of PRC (see Eq.~\eqref{eq:wf}, diamonds) is quantified by the 
normalized error of fit $\sigma$, as defined in Eq.~(\ref{eqsig}). 
Notice that in the former case, $\sigma$ is small and practically independent of $\e$; 
here $\sigma$ is due to inaccuracy of numerical differentiation of $\vp$ and of the
kernel estimation procedure.
In the latter case, the error becomes quite large for increasing $\e$, 
what cannot be explained by numerical precision only; 
large values of $\sigma$ demonstrate that
the coupling function cannot be anymore represented as 
$Q(\vp,\psi)=Z(\vp)p(\psi)$. 
(b,c) Coupling functions of the harmonically driven Rayleigh oscillator for
weak, $\e=0.01$ (panel (b)), and strong, $\e=0.55$ (panel (c)), forcing.
}
\label{fig1}
\end{figure}

First, we demonstrate that the phase description is well-defined for 
the whole explored range of $\e$. 
For this purpose we compute, for all available $\vp_k$, $\dot\vp_k$, $\psi_k=\nu t_k$,  
the rest term of our full model,
\begin{equation}
\xi_k=\xi(t_k)=\dot\vp_k-\w-Q(\vp_k,\psi_k,\e)\;,
\end{equation}
and quantify the quality of the fit by 
\begin{equation}
\sigma=\text{std}[\xi]/\text{std}[\dot\vp]\;,
\label{eqsig}
\end{equation}
where 
$\text{std}[u]=\left [ \overline{(u-\bar{u})^2} \right ]^{1/2}$
and bar denotes the time averaging over the available time series.
The dependence $\sigma(\e)$ is shown in Fig.~\ref{fig1}a, together
with a similar measure for the Winfree form~\eqref{eq:wf}. 
The latter increases with coupling strength,  
what clearly demonstrates that the Winfree form is valid for small couplings only.
In contradistinction, the full coupling function describes the dynamics with 
accuracy $\approx 3\%$ in the whole range of couplings.

In Fig.~\ref{fig1}(b,c) we also show the inferred coupling functions for weak, 
$\e=0.01$ 
(here $Q$ nearly coincides with $\e Q_1$), and strong, 
$\e=0.55$, forcing amplitudes. 
Although qualitatively similar, the coupling functions are clearly different, 
indicating importance of higher-order components $Q_2$ and $Q_3$.  

\begin{figure}[thb!]
\centerline{\includegraphics[width=0.5\textwidth]{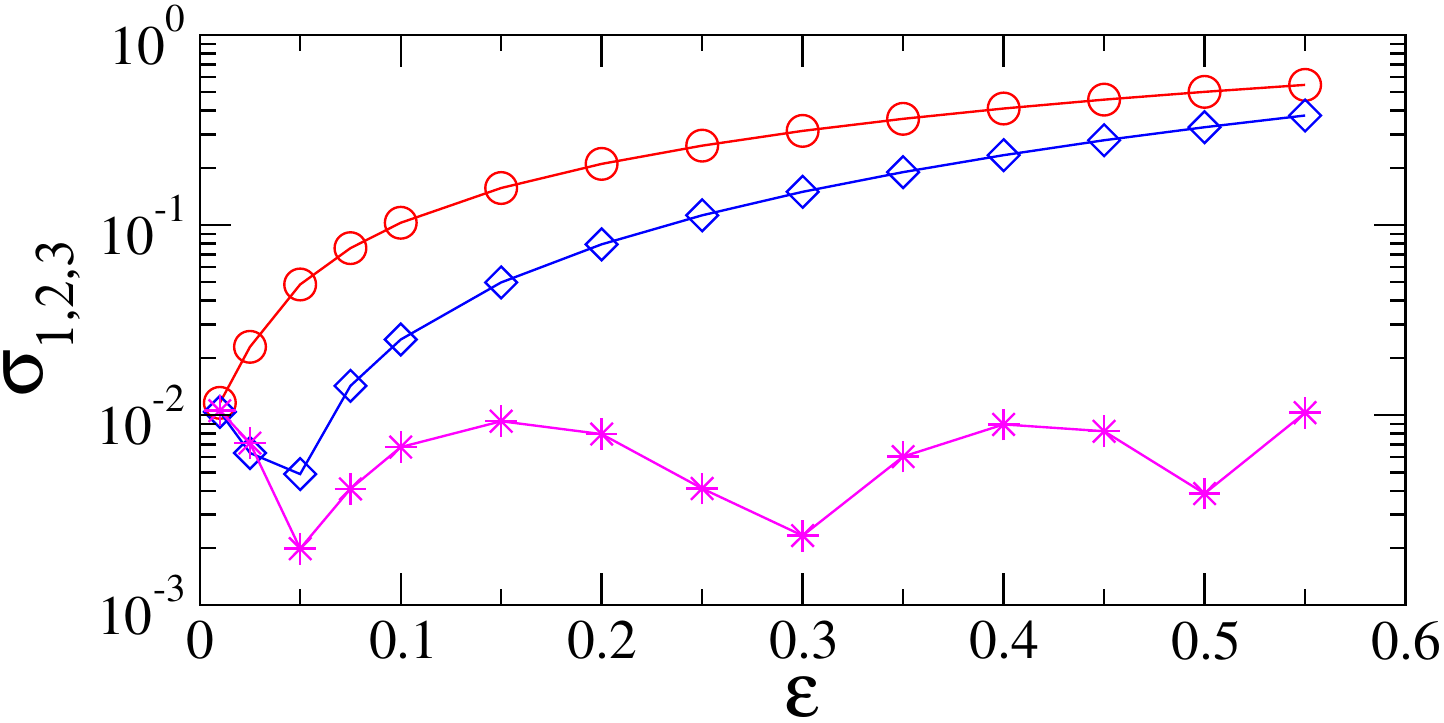}}
\caption{Error of the 1st-order (circles), 2nd-order (diamonds), and 3rd-order (stars) 
approximations of the coupling functions for the Rayleigh oscillator. 
Notice that the precision of the third-order approximation is below 1\% for the 
whole range of the driving strength. 
}
\label{fig2}
\end{figure}

\begin{figure}[thb!]
\centerline{\includegraphics[width=0.5\textwidth]{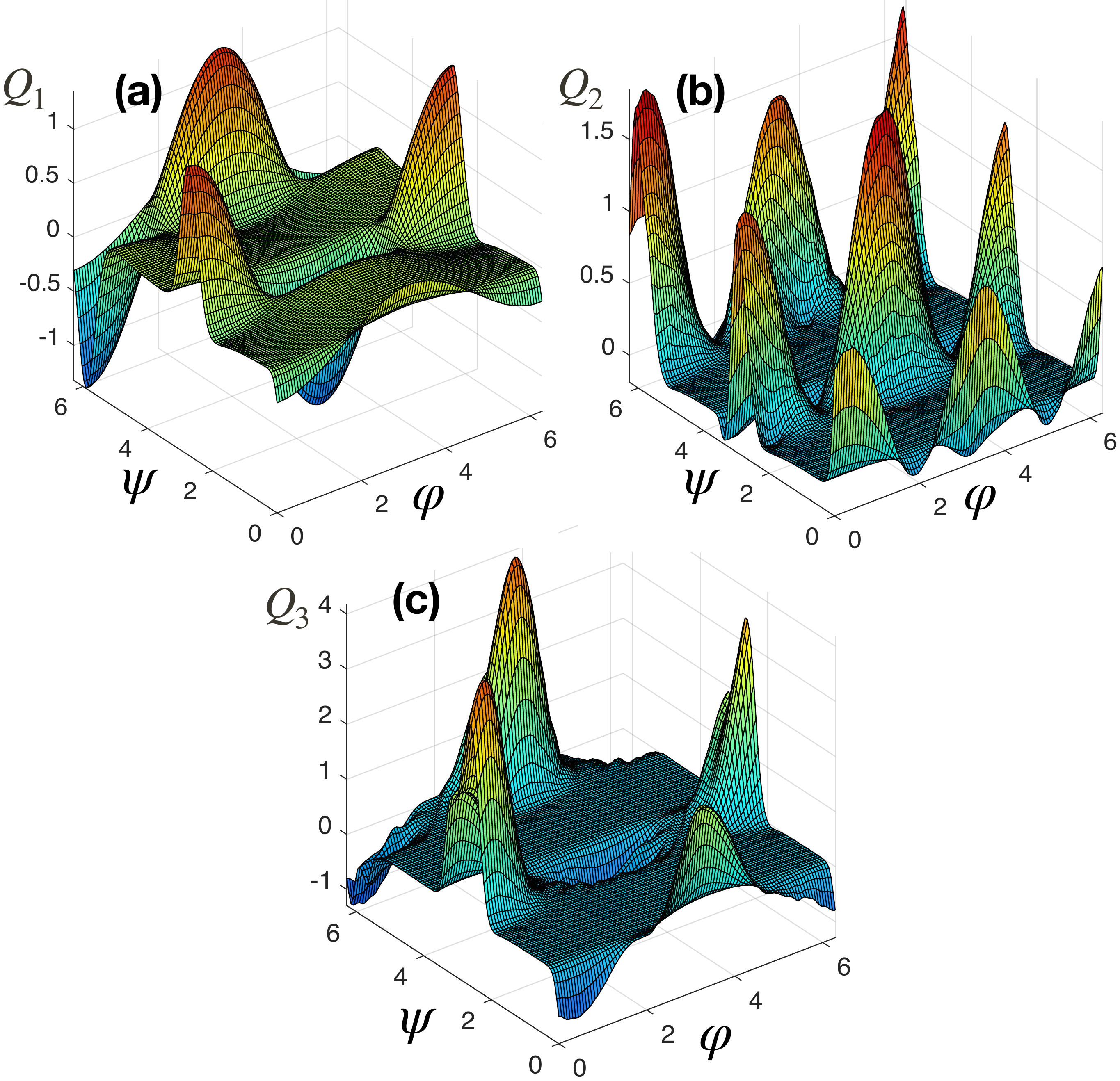}}
\caption{Coupling function of strongly perturbed Rayleigh oscillator
can be well approximated by a power series Eq.~(\ref{eq3}) with linear, quadratic, and 
qubic components, shown in (a), (b), and (c), correspondingly.
}
\label{fig3}
\end{figure}

Next, we obtain functions $Q_{1,2,3}$ as discussed in Section~\ref{secdecomp} 
and verify the validity of the series representation by Eq.~(\ref{eq3}).
To this end we compute the error of the first-, second-, and third-order approximations
\begin{equation}
\sigma_m(\e)=\text{std}\left[Q(\e)-\sum_{i=1}^m \e^i Q_i\right]/\text{std}\left[Q(\e)\right],
\label{eqsig123}
\end{equation}
where $m=1,2,3$ and $\text{std}\left[u\right]=\langle(u-\langle u \rangle)^2\rangle^{1/2}$.
Here the averaging is performed over the torus on which the coupling function
is defined:
\[
\langle w\rangle=(4\pi^2)^{-1}\int_0^{2\pi}d\vp\int_0^{2\pi}d\psi\, w(\vp,\psi)\;.
\] 
These errors are shown in Fig.~\ref{fig2}, while the 
functions $Q_{1,2,3}$ are given in Fig.~\ref{fig3}.

Fig.~\ref{fig2} shows that,
 while the first order approximation is quite accurate for $\e\lesssim 0.1$,
for larger coupling strengths the higher order terms become relevant. The third-order approximation
provides a rather good accuracy (better than 1\%)
in the whole explored range of coupling strengths. Remarkably,
the coupling functions $Q_1,Q_2,Q_3$ are very different. While the first-order coupling function $Q_1$
reproduces the known theoretical result \eqref{eq:wf}, the second-order term $Q_2$ contains strong 
second harmonics in both phases. 
Preliminary computations also show that $Q_1$ is independent of the driving frequency $\nu$, as expected,
while the higher-order functions depend on $\nu$; this issue shall be analyzed in details separately. 

\subsection{The R{\"o}ssler oscillator}
\begin{figure}[thb!]
\centerline{\includegraphics[width=0.5\textwidth]{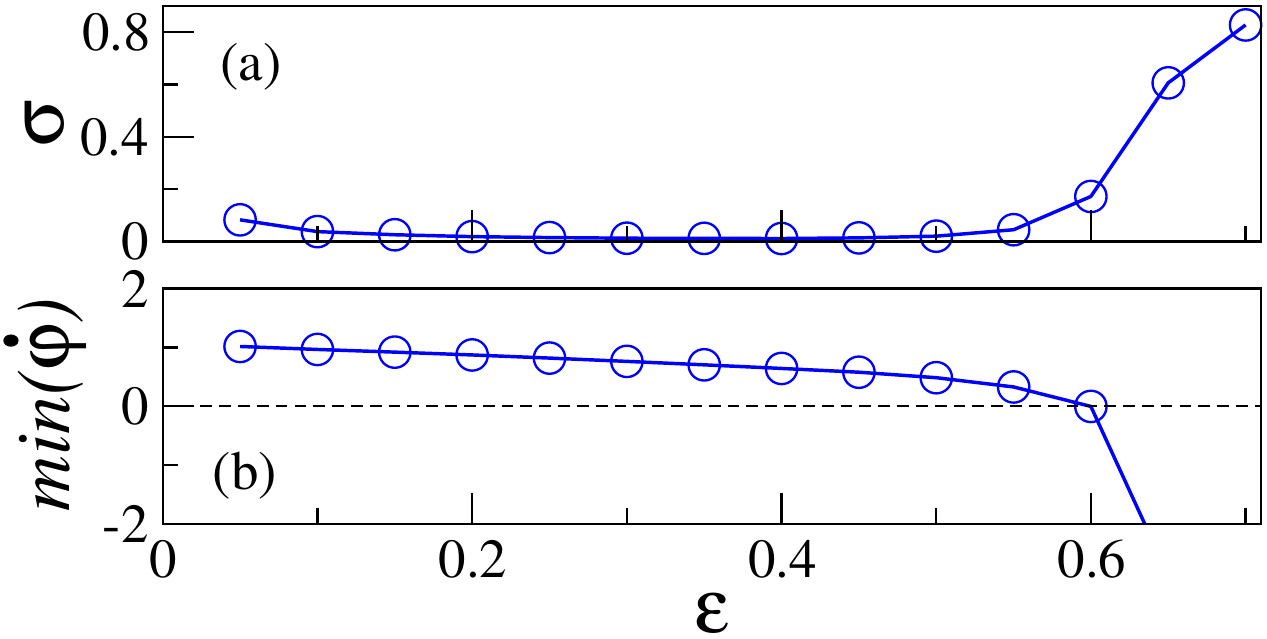}}
\caption{(a) The error of the phase model description
of the  R{\"o}ssler oscillator~(\ref{ross}) remains small for
the coupling strength $\e\lesssim 0.55$. For stronger coupling the 
description fails. This is related to the fact that the phase becomes
not monotonic, as can be seen from the plot of $\text{min}(\dot\vp)$
vs. $\e$; see text  and Fig.~\ref{fig7} for a discussion.
}
\label{fig4}
\end{figure}

\begin{figure}[thb!]
\centerline{\includegraphics[width=0.5\textwidth]{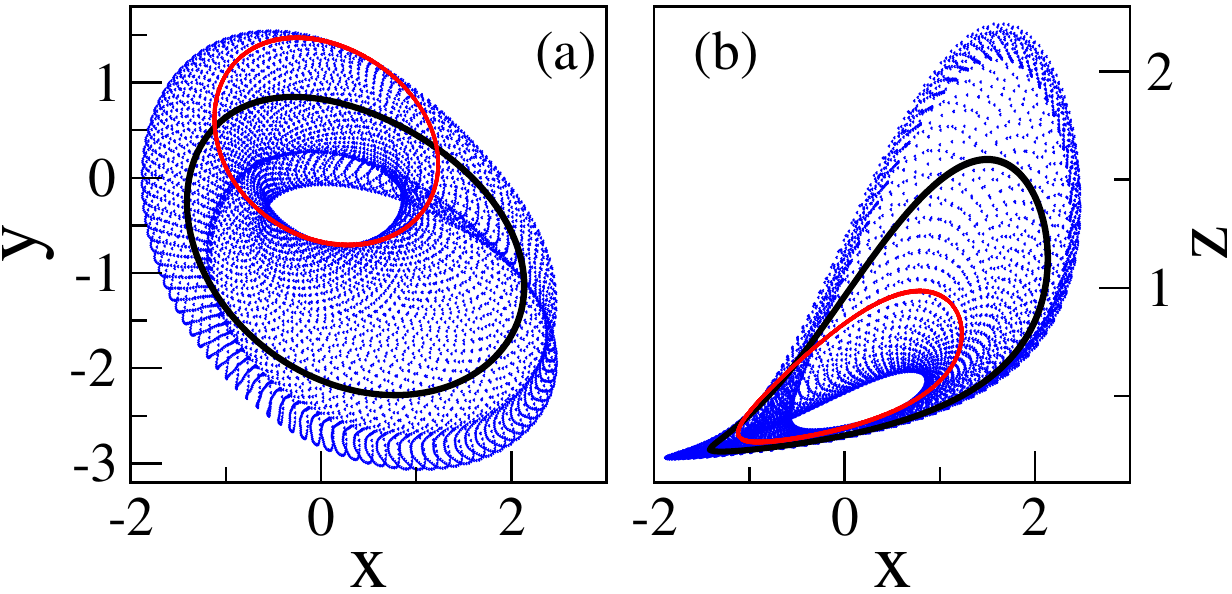}}
\caption{Two projections of the phase space of the R{\"o}ssler 
oscillator~(\ref{ross}). In both panels black bold line 
shows the unperturbed limit cycle, while the blue points show the trajectory 
of the driven system with $\e=0.55$. The red line shows 
the Poincar\'e section of the attractor obtained by fixing the phase 
of the external force; this section confirms that the trajectory of the 
strongly driven system lies on the torus.
}
\label{fig5}
\end{figure}

\begin{figure}[thb!]
\centerline{\includegraphics[width=0.5\textwidth]{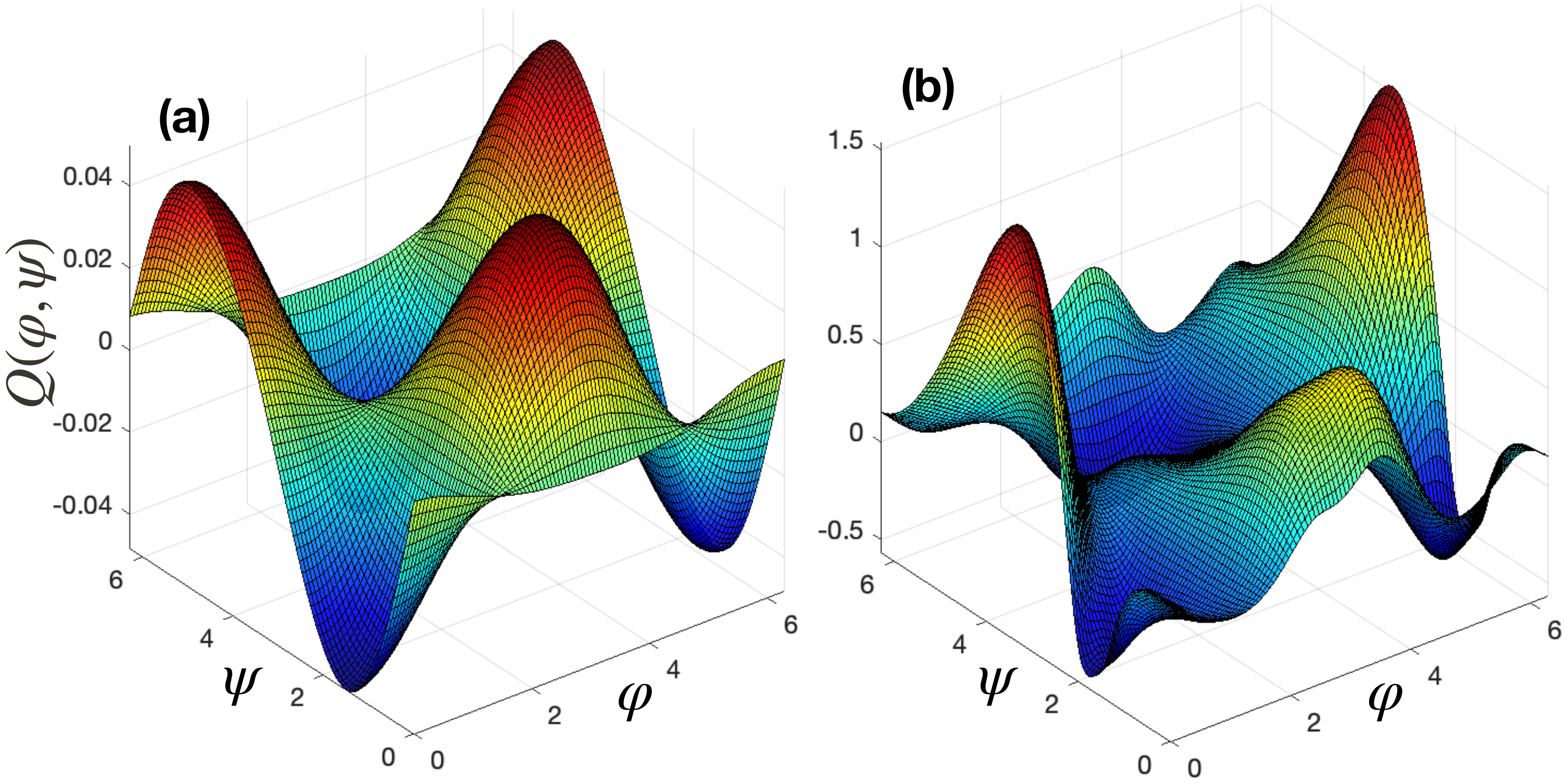}}
\caption{Coupling functions of the harmonically driven Rayleigh oscillator for
weak, $\e=0.05$ (panel (a)), and strong, $\e=0.5$ (panel (b)), forcing.
}
\label{fig6}
\end{figure}

For the second example we take a three-dimensional system,
namely the harmonically driven R{\"o}ssler oscillator:
\begin{equation}
\begin{array}{rcl}
\dot{x}&=& -y-z + \e \cos (\nu t) \;,\\
\dot{y}&=& x+0.34 y\; ,  \\
\dot{z}&=& 0.8+z(x-2) \;. 
\end{array}
\label{ross}
\end{equation}
For the chosen parameter values, the unforced system has a limit cycle
with complex multiplicators $(-8.71\pm 12.4i)\cdot 10^{-3}$.
Thus, the cycle is weakly stable; moreover, complex-valued multipliers mean 
that the dynamics cannot be reduced to a
two-dimensional one. However, our approach successfully constructs the 
phase dynamics model. To demonstrate this, we infer the coupling 
functions for $\nu=0.5$ and forcing strengths varied from $\e=0.05$ to $0.7$.
The quality of the model for different $\e$ is illustrated in Fig.~\ref{fig4}, where
we show the normalized standard deviation $\sigma(\e)$, defined according to~\eqref{eqsig}.
One can see that the model remains valid for the coupling values as large as 
$\e=0.55$. We emphasize, that for such a driving the deviation of the trajectory
from the unperturbed limit cycle is not small at all, see Fig.~\ref{fig5}, where the 
attracting torus at this regime is depicted. For the coupling functions for weak 
and strong coupling see Fig.~\ref{fig6}. Qualitatively, the results
are similar to that for the Rayleigh oscillator: The coupling function is strongly
nonlinear, as one can conclude from the comparison of two panels of Fig.~\ref{fig6}.

\begin{figure}[thb!]
\centerline{\includegraphics[width=0.5\textwidth]{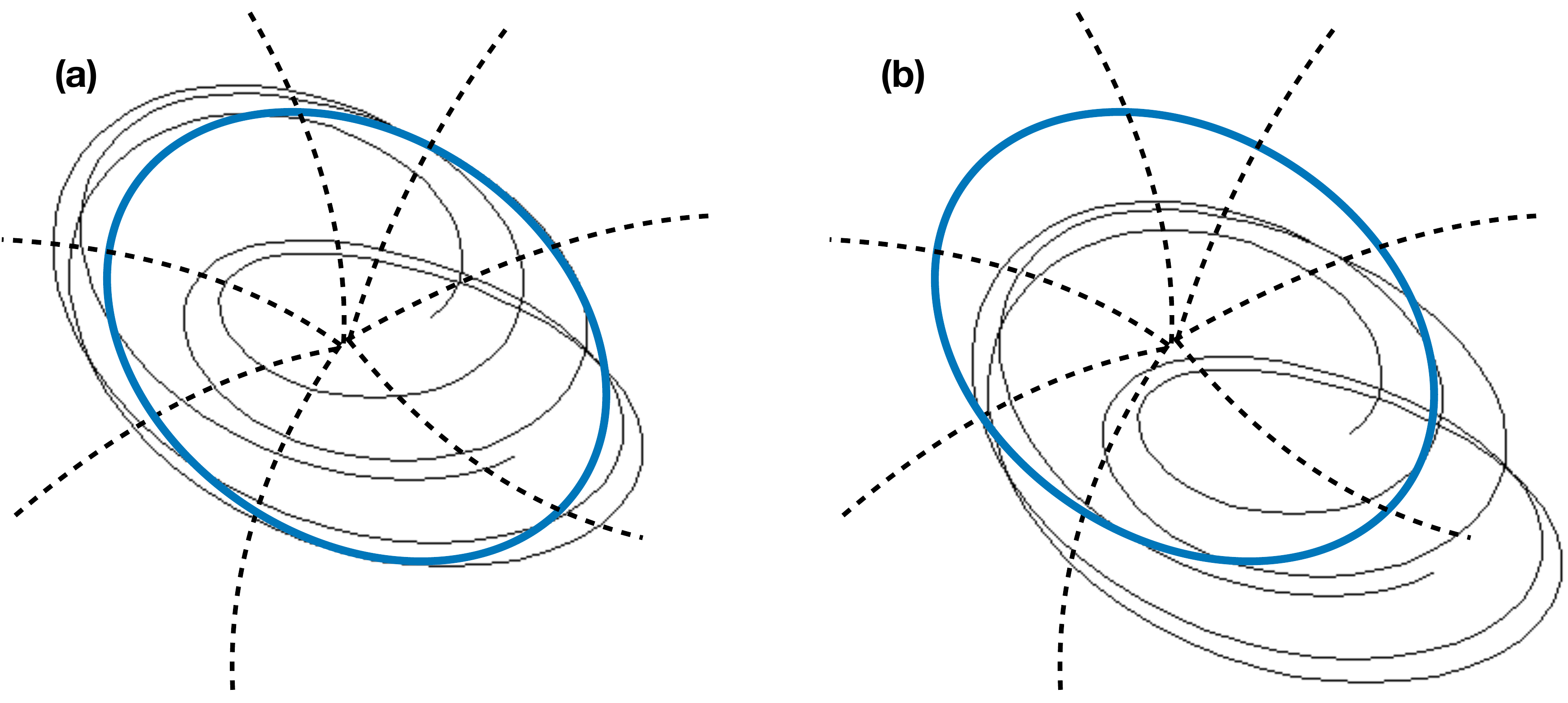}}
\caption{Qualitative illustration of a limitation of the phase 
reduction technique. Here, the bold blue line presents the 
unperturbed limit cycle, while the black solid curve shows a piece
of the trajectory in the phase space.
In (a) the loops of the trajectory revolve around the
origin. The point within one revolution goes sequentially through all the isochrons (dashed lines)
and the phase therefore grows monotonically. In (b) several loops do not revolve around
the origin, what means that two points within each such loop have same phase
(this also means that the phase is not monotonically growing).
Thus, the definition of a cycle becomes ambiguous and the phase reduction fails.   
}
\label{fig7}
\end{figure}

We now discuss, why the technique fails for larger coupling, $\e\gtrsim 0.6$.
In the case of the Rayleigh oscillator, the value of $\e$ was limited  by 
the effect of locking to the drive, what makes the inference of the 
coupling function impossible. In  the case of the R{\"o}ssler system, the reason
for failure is different. Here, the torus of the driven system in the phase space 
becomes strongly shifted with respect to the original cycle, so that the
trajectory starts to cross "wrong" isochrons, see Fig.~\ref{fig7}. In other
words, the isochrons do not anymore represent a proper phase coordinate on the 
attracting torus.
This results in a non-monotonic growth of the  isochron-based phase, manifested by the 
violation of the ``good-phase-condition'' $\dot\vp>0$, cf. Fig.~\ref{fig4}b.
   
\section{Discussion}

In summary, we have developed a simple and efficient technique for numerical 
determination of the phase of a perturbed self-sustained oscillator from the 
equations formulated in state space variables. 
In its turn, computation of 
the phase provides an easy way to reconstruct the equation (\ref{eq3}) for 
the phase dynamics, where the coupling function is given on an equidistant 
grid within $0\le \vp<2\pi, 0\le \psi<2\pi$ or/and can be given by a 
double Fourier series in phases $\vp,\psi$. 
For transparency of the presentation we have considered 
driven systems, but extension to the case of two coupled oscillators is 
straightforward. The phases can be in the same way obtained for a network of more 
than two oscillators as well, but reconstruction of high-dimensional coupling 
functions requires extremely long data sets and becomes unfeasible, 
see a discussion in Ref.~\onlinecite{Kralemann-Pikovsky-Rosenblum-14}. 

We stress here, that our approach heavily relies on the availability 
of full dynamical equations, as they are needed to determine the true phase based 
on the isochrons. The existing approaches for phase reconstruction from a scalar time series
\cite{Kralemann_et_al-07,Kralemann_et_al-08,Kralemann_et_al-13,Kralemann-Pikovsky-Rosenblum-14}
do not provide the exact isochron-based phase, what appears
to be crucial for the proper reconstruction of the coupling function. 
Development of advanced phase reconstruction methods is a subject of 
ongoing research and our approach provides an easy way to test their efficiency. 

We have demonstrated that reduction to the phase description is valid even
for quite strong driving/coupling and have confirmed numerically 
that the coupling function can be represented as a power series of 
the driving strength.
The suggested algorithm for correct computation of the phase offers also 
an easy way to obtain the PRC of the system. However, our simulations show that 
description in terms of an (amplitude-dependent) PRC fails for strong coupling:
one cannot generalize the Winfree-type coupling~\eqref{eq:wf} to the full
coupling function as $Q(\vp,\psi,\e)=Z(\vp,\e)p(\psi)$.  
This fact has strong implications for the 
time-series analysis of experimental data: if only data for a particular 
forcing amplitude are available, then, to be safe, one should assume a nonlinear 
regime and reconstruct the coupling function in its general form, $Q(\vp,\psi)$, 
but not in the Winfree representation.
Thus, one does not have to rely on the assumption of week coupling, that can 
hardly be checked if only passive observation of oscillatory 
system is possible.
Furthermore, if data for several amplitudes of forcing are available, 
nonlinearity level could be potentially revealed from such an experiment.

We have illustrated two effects that limit applicability of the approach. 
The first one is synchronization of the system 
to the external perturbation, as in the case of the Rayleigh oscillator.  
In the example of the Roessler system, the technique does not work if a 
strong perturbation makes the 
mapping $\Phi(\mathbf{X})$ not unique within one cycle. 
Finally, we expect that the approach can also fail if a very strong coupling 
destroys the smooth torus in the phase space,
see \citep{Afraimovich-Shilnikov-83} for a rigorous treatment 
and \cite{Pikovsky-Rosenblum-Kurths-01} for a qualitative discussion
of this mechanism.

Our results open a way for further studies on phase reduction, e.g. 
they provide a numerical tool for an analysis of the dependence 
of high-order terms of the coupling functions on parameters of the 
driving, with the utmost goal to extend the analytical techniques.
Finally, our approach can be used as a benchmark for testing technique for coupling 
functions reconstruction from data 
\cite{Kralemann_et_al-07,Kralemann_et_al-08,Kralemann_et_al-13,PhysRevLett.109.024101}
where only one observable of the system is available. 
  
\acknowledgments
The work was supported by ITN COSMOS (funded by the European Union Horizon 2020 
research and 
innovation programme under the Marie Sklodowska-Curie grant agreement No. 642563). 
Numerical part of this work was supported by the Russian Science
Foundation (Rayleigh model: Grant No. 17-12-01534; Roessler model: Grant No. 14-12-00811).

%

\end{document}